\documentclass[pra,twocolumn,showpacs,superscriptaddress]{revtex4-1}
\usepackage{color}
\usepackage{times}
\usepackage{amsmath}
\usepackage{bm}
\usepackage{graphicx} 
\usepackage{multirow}
\usepackage{hyperref}

 \newcommand{\eref}[1]{(\ref{#1})} %short form
\newcommand{\eeref}[1]{Eq.~(\ref{#1})} %full form
\newcommand{\fref}[1]{Fig.~\ref{#1}}

\newcommand{\sref}[1]{section \ref{#1}}

\newcommand{\mbf}[1]{\mathbf{#1}}

\newcommand{\x}{\mathbf{x}}

\newcommand{\eqn}[1]{\begin{eqnarray}#1 \end{eqnarray}}

\newcommand{\ecut}{\epsilon_{\rm{cut}}}

\newcommand{\EQ}[1]{\begin{eqnarray}#1\end{eqnarray}}

\newcommand{\rC}{\textbf{C}} % Classical region
\newcommand{\rI}{\textbf{I}} % Quantum region
\newcommand{\PC}{\mathcal{P}_{\rC}}

\def\x{\mathbf{x}}
\def\y{\mathbf{y}}

\definecolor{MyGray}{rgb}{.2,.5,.3}

\begin{document}

\title{Suppression of Kelvon-induced decay of quantized vortices in oblate Bose-Einstein Condensates}
\author{S. J. Rooney} 
\author{P. B. Blakie} 
\affiliation{Jack Dodd Center for Quantum Technology, Department of Physics, University of Otago, Dunedin, New Zealand.}
\author{B. P. Anderson}
\affiliation{College of Optical Sciences, University of Arizona, Tucson, Arizona 85721, USA}
\author{A.~S. Bradley} 
\affiliation{Jack Dodd Center for Quantum Technology, Department of Physics, University of Otago, Dunedin, New Zealand.}
\date{\today}
\begin{abstract}
We study the Kelvin mode excitations on a vortex line in a three-dimensional trapped Bose-Einstein condensate at finite temperature. 
Our stochastic Gross-Pitaevskii simulations show that the activation of these modes can be suppressed by tightening the confinement along the direction of the vortex line, leading to a strong suppression in the vortex decay rate as the system enters a regime of two-dimensional vortex dynamics. 
As the system approaches the condensation transition temperature we find that the vortex decay rate is strongly sensitive to dimensionality and temperature, observing a large enhancement for quasi-two-dimensional traps. Three-dimensional simulations of the recent vortex dipole decay experiment of Neely {\em et al.} [Phys. Rev. Lett. \textbf{104}, 160401 (2010)] confirm two-dimensional vortex dynamics, and predict a dipole lifetime consistent with experimental observations and suppression of Kelvon-induced vortex decay in highly oblate condensates.
\end{abstract}
\maketitle
%============================================================================
\section{Introduction}
Quantum vortices in Bose-Einstein condensates (BECs) are topologically stable excitations of the superfluid matter wave field that carry angular momentum~\cite{Fetter2001,*Fetter09a}. Unless the system is rotating sufficiently fast, vortices are thermodynamically unstable~\cite{Rokhsar97a}.  At finite temperatures the thermal cloud can provide a source of dissipation and noise, and hence allow vortex decay. However, there is currently little understanding of the decay mechanism and its timescale. The dynamics of vortices in such systems is therefore of interest as a robust test of non-equilibrium field theory. A number of different methods have been developed to describe the dynamics of vortices in finite-temperature systems. Single vortex decay has been simulated using the Zaremba-Nikuni-Griffin formalism \cite{Jackson09a}, the classical field method \cite{Schmidt2003}, and within a Gross-Pitaevskii theory with phenomenological damping~\cite{Madarassy08a}. The nature of the thermal instability has also been studied within a variational approach~\cite{Duine2004} to the stochastic Gross-Pitaevskii equation~\cite{Stoof1999,*Stoof2001}.  

Classical vortices can support long-wavelength helical traveling waves known as Kelvin waves~\cite{Thompson1880}. Superfluid Kelvin waves, or Kelvons~\cite{Fetter04a}, have long be studied in Helium~\cite{Tsubota09a}, and more recently have become a topic of interest in BECs. The Kelvon excitation mechanism has been theoretically investigated \cite{Mizushima03a}, along with the microscopic dynamics of coherently excited Kelvin waves~\cite{Simula08a,*Simula08b}, with results consistent with the experimental evidence for Kelvons \cite{Bretin03a}.  Similar bending waves have also been shown to play an important role in the instability of vortex rings in trapped BECs \cite{Horng2006a}.

\begin{figure}[!t]
\begin{center}
\includegraphics[width=\columnwidth]{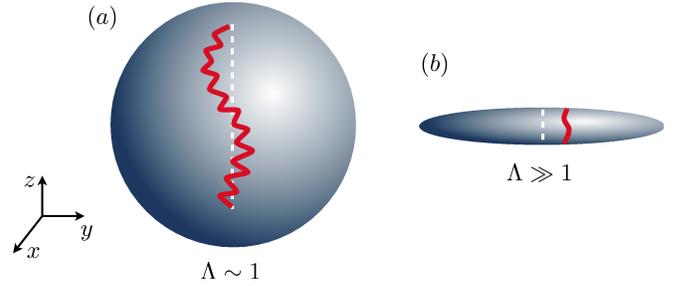}
\caption{Schematic of vortex (solid line) bending from the $z$-axis (dashed line) in a trapped BEC for (a) a spherical trap  (in terms of trap frequencies defined in Eq.~\ref{Vdef}, $\Lambda=\omega_z/\omega_r\sim1$), and (b) a highly oblate trap ($\Lambda\gg1$).  
\label{fig1}}
\end{center}
\end{figure}

While the coherent excitation of Kelvin modes in a BEC is well understood, their role in the finite-temperature system is largely unexplored. Here we show that the thermal activation of Kelvin waves is a dominant factor in the decay of a vortex for a finite-temperature three-dimensional (3D) BEC. The underlying reason is that Kelvin waves cause the vortex to wobble and emit acoustic radiation, thus allowing the vortex to more effectively dissipate energy, and hence decay by moving out to the boundary of the BEC. As Kelvin waves are fundamentally 3D excitations of the vortex line, they may be suppressed by flattening the system  (see Fig.~\ref{fig1}). 

Indeed, beyond a certain oblateness we find that the Kelvon mode contribution is essentially frozen out and the vortex decay rate reduces to a geometrically-invariant value, characterizing a regime of two-dimensional vortex (2DV) dynamics. We also find that for tighter traps and higher temperatures the matter wave itself crosses over to being quasi-2D and we find a dynamical signature of the phase-fluctuating condensate: an anomalous increase in the rate of vortex decay. 

As a further application of these concepts, we model a recent experiment~\cite{Neely10a} studying vortex dipoles in the 2DV regime, confirming the 2DV behavior in finite temperature dynamics. The decay characteristics are found to be qualitatively very different from 3D dipole decay which exhibits vortex ring and loop formation, accelerating the damping process.
 
Our theoretical framework for studying this system is the stochastic projected Gross-Pitaevskii equation (SPGPE)~\cite{SGPEI,*SGPEII}, which is a grand-canonical c-field method~\cite{Blakie08a} (also see  \cite{Proukakis2006a,*Proukakis2006b,*Proukakis2008a,Nestor2011a} for related theories).  This formalism has been used to study vortex lattice formation from a rotating thermal cloud \cite{Bradley08a}, a quantitative model of spontaneous vortex formation in experiments~\cite{Bradley08a,Weiler08a}, and the decay of a single vortex \cite{Rooney10a} (also see \cite{Wright09a,*Wright08a,*Wright10a}).   

%=====================================================================================
 %=====================================================================================

%=====================================================================================
\begin{figure*}[!t]
\begin{center}
\includegraphics[width=\textwidth]{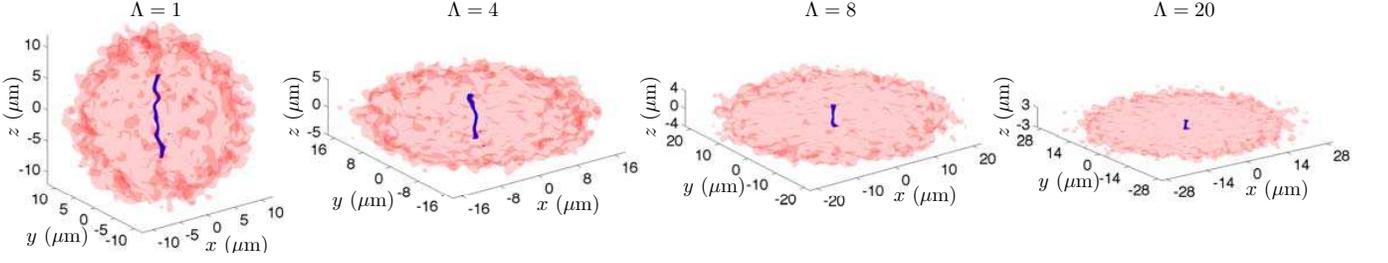}
\caption{Density isosurfaces of the $\rC$-field after $t = 0.61~{\rm{s}}$ of SPGPE evolution, for a range of trap geometries. Red: the full density isosurface. Blue: isosurface restricted to the region of the vortex core. The system temperature is $T = 0.78T_c^0$.}
\label{figiso}
\end{center}
\end{figure*}
%=====================================================================================
\section{System and equation of motion}
\subsection{Physical system}
We begin by describing the system of bosonic atoms in the cold-collision regime with the second-quantized many body Hamiltonian
\EQ{\label{fullHamil}
H&=&\int d^3{x}\;\Psi^\dag(\x) H_{\rm sp}\Psi(\x)\nonumber\\
&&+\frac{g}{2}\int d^3{x}\;\Psi(\x)^\dag\Psi(\x)^\dag\Psi(\x) \Psi(\x) 
}
where the single-particle Hamiltonian is
\EQ{
H_{\rm sp}=-\frac{\hbar^2\nabla^2}{2m}+V(\mbf{x}),
}
the cylindrically symmetric harmonic trap is
\EQ{\label{Vdef}
V(\mbf{x}) = \frac{m}{2}\left[\omega_r^2(x^2+y^2)+\omega_z^2z^2\right],
} 
and $g=4\pi\hbar^2a/m$, with $m$ the atomic mass and $a$ the s-wave scattering length. The field operators obey Bose commutation relations
\EQ{
[\Psi(\x),\Psi^\dag(\y)]=\delta(\x-\y).
}
We define $\Lambda\equiv\omega_z/\omega_r$ as the degree of trap oblateness, with
$\omega_r = \Lambda^{-1/3} \bar{\omega},$
enforcing a constant geometric mean frequency $\bar{\omega}^3\equiv \omega_r^2\omega_z$ for arbitrary oblateness. This choice ensures the ideal gas transition temperature $T_c^0$ is invariant under scaling of $\Lambda$ at fixed total atom number $N$.

\subsection{Stochastic PGPE Formalism}
In this work we use the SPGPE formalism to describe the process of vortex decay in a finite temperature BEC.
The approach we follow in choosing parameters and preparing initial states for the SPGPE is extensively detailed in Ref.~\cite{Rooney10a}, hence we only briefly summarize it here.  

The SPGPE is a c-field method where the system is decomposed in terms of the single-particle modes of the system satisfying $H_{\rm sp}\phi_n(\x)=\epsilon_n\phi(\x)$, where $n$ represents all quantum numbers required to specify the modes.
The system is divided into the coherent region ($\rC$) consisting of all states with  $\epsilon_n\leq \ecut$, and the incoherent region ($\rI$) containing the remaining high energy modes \cite{Blakie08a}. This division is made  
using a projection operator, $\PC$ which defines the $\rC$-region field operator as
\EQ{
\psi(\x)=\PC\Psi(\x)\equiv \sum_{\epsilon_n\leq \ecut} \phi_n(\x)\int d^3y\;\phi_n^*(\y)\Psi(\y).
}
Equivalently, we have the expansion 
\EQ{\label{fieldOp}
\psi(\x)=\sum_{\epsilon_n\leq \ecut} a_n\phi_n(\x),
}
where $[a_n,a_m^\dag]=\delta_{n,m}$.
The $\rI$-region acts as a thermal reservoir for the $\rC$-region, and is assumed to be in thermal equilibrium at a temperature $T$ and chemical potential $\mu$.  The  $\rC$-region dynamics are described by a stochastic differential equation for the c-field corresponding to \eref{fieldOp}: 
\EQ{\label{cfield}
\psi_\rC(\x)&\equiv& \sum_{\epsilon_n\leq \ecut}\alpha_n\phi_n(\x)
}
where $\alpha_n$ are ordinary c-numbers.
The equation of motion is known as the simple-growth SPGPE~\cite{Bradley08a} 
\begin{equation}
d\psi_\rC=\PC\Big\{ -\frac{i}{\hbar}L_\rC\psi_\rC dt 
+\frac{\gamma}{k_BT}(\mu-L_\rC)\psi_\rC dt+dW\Big\},
\label{SGPEsimp}
\end{equation}
where $L_\rC$ is the Hamiltonian evolution operator for the $\rC$ region $L_\rC\psi_\rC \equiv \left(H_{\rm sp}+g|\psi_\rC|^2\right)\psi_\rC$, and the complex Gaussian noise, $dW$, satisfies $\langle dW^*(\mbf{x},t)dW(\mbf{x}^\prime,t)\rangle  =2\gamma\delta(\mbf{x}-\mbf{x}^\prime)dt,$ 
$\langle dW(\mbf{x},t)dW(\mbf{x}^\prime,t)\rangle =0$. For a full derivation of the SPGPE we refer the reader to \cite{SGPEI,*SGPEII}.  

The first term of the SPGPE (\ref{SGPEsimp}) describes Gross-Pitaevskii evolution over the $\rC$ region, while the second and third terms account for scattering of two $\rI$-region atoms, resulting in growth of the $\rC$-region, and the corresponding time-reversed process. The growth rate, $\gamma$, can be calculated in near-equilibrium systems \cite{Bradley08a} and is independent of position over a large region of the trap with the value
\begin{eqnarray}\label{gamdef} \gamma&=&\frac{4m(ak_BT)^2}{\pi\hbar^3}\sum_{j=1}^\infty\; \frac{e^{\beta\mu(j+1)}}{e^{2\beta\ecut j}}\Phi\left[\frac{e^{\beta\mu}}{e^{\beta\ecut}},1,j\right]^2,
\end{eqnarray}
where $\beta=1/k_BT$ and $\Phi$ is the Lerch transcendent.

To obtain initial states for evolution according to (\ref{SGPEsimp}), we follow the procedure used in \cite{Rooney10a}, which we briefly summarize.  We generate (non-vortex) finite-temperature equilibrium states by evolving the SPGPE with an appropriate choice of $T$, $\mu$, and $\ecut$\footnote{We require  $N$ to be fixed and the highest energy mode in  $\rC$   to be appreciably occupied ($\approx 2 $ atoms).}.
The values of $T$, $\mu$, and $\ecut$ then determine $\gamma$ [Eq.~(\ref{gamdef})] and hence all SPGPE parameters are obtained in a physically consistent manner prior to simulation.
To create the vortex initial state we imprint a phase $\psi_\rC(\x)\to \psi_\rC(\x)e^{i\Theta(\x)}$ onto the equilibrium state, where
$\Theta(\x)$ is the azimuthal angle in the radial plane~\footnote{We do not find it necessary to imprint a vortex density profile because at the temperatures under consideration the extra energy of the core with phase-only imprinting is negligible compared with that from thermal fluctuations}. This creates a straight singly-charged vortex line along the $z$ axis of the trap. This procedure generates a non-equilibrium state because the reservoir parameters in (\ref{SGPEsimp}) are for a non-rotating reservoir. 

%====================================================
\begin{figure}[!t]
\begin{center}
\includegraphics[width=0.9\columnwidth]{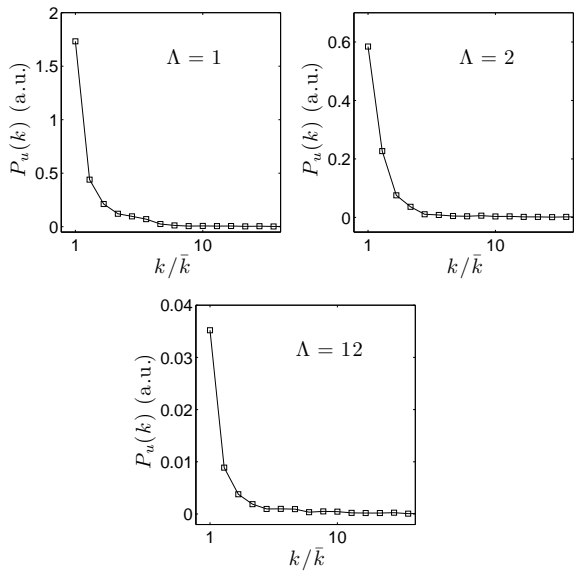}

\caption{Power spectra for a range of geometries as a function of $k_z / \bar{k}$, where $\bar{k} = \pi /\mathcal{W}$ corresponds to the largest wavelength resolvable for the chosen window $\mathcal{W}=R^z_{TF}/2$ .  The spectra are calculated after $t = 0.70$~s, and for $T=0.78T_c^0$.
\label{figspectra}}
\end{center}
\end{figure}
%====================================================
\section{Single Vortex Decay \label{results}}
In our simulations we use initial vortex states prepared for a system containing $5\times 10^5$ $^{87}$Rb atoms in a trap of (fixed) geometric mean trap frequency $\bar{\omega}=2\pi\times 19.7{\rm s^{-1}}$ [for which  $T_{\rm{c}}^0 = 69.7$nK], and a range of geometries from spherical to highly oblate.  For each parameter set we calculate vortex decay properties by evolving 10-30 trajectories of the SPGPE~\footnote{The number of trajectories required to obtain good statistics increases with the system temperature}.

Fig.~\ref{figiso} shows density isosurfaces of the vortex in a variety of condensate geometries after $t = 0.61 ~{\rm{s}}$ of evolution.  The phase imprinting technique leads to a straight vortex line along the $z$-axis, but under evolution the vortex rapidly thermalizes, typically within $0.1 ~{\rm{s}}$ (see discussion below).   We observe a significant amount of vortex bending in the spherical geometry and as the condensate becomes more two-dimensional,  the magnitude of bending lessens and the vortex behaves essentially as a straight line.  Although vortex bending on a short length scale is evident in spherical geometries, the vortex stays approximately aligned with the $z$-axis throughout its evolution.

\subsection{Kelvon power spectrum}
To obtain the vortex line coordinates during evolution we locate the radial position of the phase singularity $(x_v, y_v)$ in each plane (i.e.~$z$ value) and for all times $t$.  Fluctuations in the phase at the condensate boundary restrict the spatial range over which the vortex can be detected. This reduces the range of $z$-values for which we can obtain the vortex-line coordinates, and, as time evolves and the vortex precesses toward the radial condensate boundary, this range of $z$-values further reduces. For this reason we limit vortex detection to the range $z = \pm \mathcal{W}$, where $\mathcal{W}=\frac{1}{2} R^z_{{\rm{TF}}}$, with $R^z_{{\rm{TF}}}$ the Thomas-Fermi radius along the $z$ direction. We have found this to be the largest range which gives a well defined and unique curl signal over the vortex lifetime for our simulations. 
   
We define  $u(z,t) = [x_v(z,t) - \bar{x}(t)] + i[y_v(z,t) - \bar{y}(t)]$ to be the (complex) vortex line displacement from its instantaneous mean, where the barred quantities are the vortex positions averaged over $z$.  To quantify the extent of vortex bending at a given time, we take the Fourier transform with respect to $z$ (denoted by $\mathcal{F}_z$) of $u(z,t)$, and calculate the power spectrum
 \begin{equation}
 P_u(k,t)=| \mathcal{F}_z [ u(z,t) ] | ^2.
 \end{equation}
 Note that this measure of the power spectrum is insensitive to varicose waves~\cite{Simula08b} and is thus entirely due to vortex bending. 
 
 %====================================================
%TOTAL POWER
\begin{figure}[!t]
\begin{center}
\includegraphics[width=0.8\columnwidth]{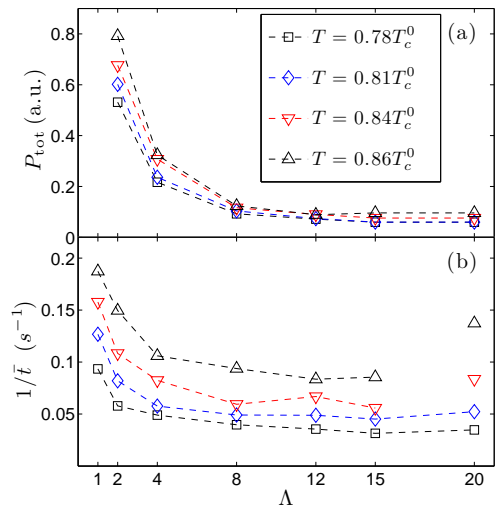}
\caption{(a) Total power as a function of trap oblateness $\Lambda$. (b) Effective vortex decay rate ($1/\bar{t}$) as a function of $\Lambda$. The disconnected points for $\Lambda=20$ break from the trend of vortex decay rate independent of $\Lambda$, as discussed in \sref{anomdecay}.}
\label{figratpow}
\end{center}
\end{figure}
%====================================================
Fig.~\ref{figspectra} shows the power spectra vs Kelvon wavenumber for a range of geometries at $t = 0.70$~s. We rescale the $k$~axis by $\bar{k} =   \pi / \mathcal{W}$, which is the wavenumber corresponding to the largest visible wavelength within our choice of window length.  In all geometries we see the power spectra take a thermalized form, with maximum occupation of the long-wavelength bending modes (low values of $k$).  
 The thermalization of the vortex line from the straight initial conditions is most directly monitored using the Kelvon power spectrum. We find that typically within $\sim100$ ms the Kelvon spectrum reaches a stable state. This is much shorter than the vortex decay time (see Sec.~\ref{SecEffOfGeom}), so we can rule out the influence of any initial state artifacts on our predictions for the vortex decay process. 
 
 %====================================================
\begin{figure}[!t]
\begin{center}
\includegraphics[width=0.95\columnwidth]{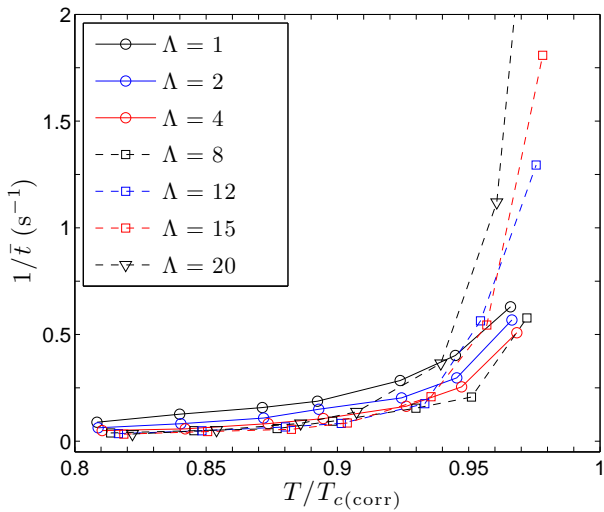}
\caption{Vortex decay rate as a function of relative temperature, for a range of geometries.  Here the critical temperature is adjusted to account for interactions and finite size effects (\eeref{Tccorr}). The highest decay rate (for $\Lambda = 20$) is at $T/T_{c{\rm{(corr)}}} = 0.98$, giving $1/\bar{t} = 3.9\; {\rm s}^{-1}$. 
\label{figvel}}
\end{center}
\end{figure}
%====================================================

\subsection{Effect of geometry: Kelvon power and vortex decay}\label{SecEffOfGeom}
 The effect of geometry on the importance of the Kelvon modes is clearly visible comparing  Figs.~\ref{figspectra} (a)-(c): The total power in the spectrum decreases significantly as the system becomes more oblate.
It is useful to calculate this total power in the bending modes, $P_{\rm{tot}}= \int P_u(k,t) {\rm{d}}k$, which by Parseval's theorem is related to the line-averaged deviation of the vortex from its mean position~\footnote{To obtain good statistics we integrate an instantaneous $P_u(k)$ curve (as in Fig.~\ref{figspectra}), and then time-average the result for $0.5{\rm s}$. This time is chosen to be much shorter than the vortex decay-time (which exceeds $5{\rm s}$ for all parameters), to give a short-time ensemble-average of the power data.}. Results for the total power are shown in Fig.~\ref{figratpow} (a). It decreases exponentially as the geometry becomes more two-dimensional, and we see that the total power is independent of oblateness for the most highly oblate condensates. We see that vortex bending is significantly reduced in the highly oblate systems, being almost entirely absent for $8<\Lambda$.

Fig.~\ref{figratpow}(b) shows the effect of condensate geometry on the vortex lifetime, for a range of system temperatures. The lifetime is quantified here in terms of the \emph{mean first exit time}~\cite{Rooney10a}, denoted by $\bar{t}$. 
This is calculated as the ensemble average of the time it takes for the vortex to become indistinguishable from thermal fluctuations at the BEC edge (in the plane $z = 0$). The vortex decay rate, $1/\bar{t}$, decreases with increasing $\Lambda$, following the same trend as the total bending mode power [\fref{figratpow}(a)]. However, rather than decreasing to zero, the decay rate approaches a constant value of order $\sim 0.1{\rm s}^{-1}$, being almost independent of $\Lambda$ for  $8< \Lambda < 20$.
This constant is determined by purely 2D vortex dissipation processes which do not involve bending. Interestingly, for the highest temperatures considered, the decay rate starts to increase again at $\Lambda=20$. This increase is not associated with an increase of bending mode power [\fref{figratpow}(a)]. Aside from this anomaly, discussed below, we infer a critical oblateness of $\Lambda_c\approx 8$ for the onset of 2DV dynamics in our system.

\subsection{Effect of temperature: Anomalous decay in quasi-2D systems  } \label{anomdecay}

The results of the previous section show that the vortex enters a 2D dynamics regime at $\Lambda\sim 8$, associated with strong suppression of bending modes, and a diminished vortex decay rate. However, for the highest temperatures the decay rates exhibit a revival as $\Lambda$ increases. We interpret this as arising from the system entering a phase fluctuating quasi-2D regime.

First, to more accurately characterize the critical temperature we include the  downward shifts from $T_c^0$ due to interaction and finite-size effects \cite{Giorgini96a}. This gives a corrected critical temperature
\eqn{T_{c({\rm{corr}})} = T^0_c + \delta T^{\rm fs}_c + \delta T^{\rm{int}}_c  ,\label{Tccorr}}
where $\delta T^{\rm fs}_c= -0.73 \frac{\omega}{\bar{\omega}} N^{-1/3} T_c^0$ is the first-order correction due to finite-size effects, and $\delta T^{\rm{int}}_c = -1.33 \frac{a}{\bar{a}} N^{1/6} T_c^0$ is the first order correction due to interactions, with $\omega = (2\omega_r + \omega_z)/3$ and $\bar{a} = (\hbar /m \bar{\omega})^{1/2}$. 
%%====================================================
%isosurfaces
\begin{figure*}[t!]
\begin{center}
\includegraphics[width=0.85\textwidth]{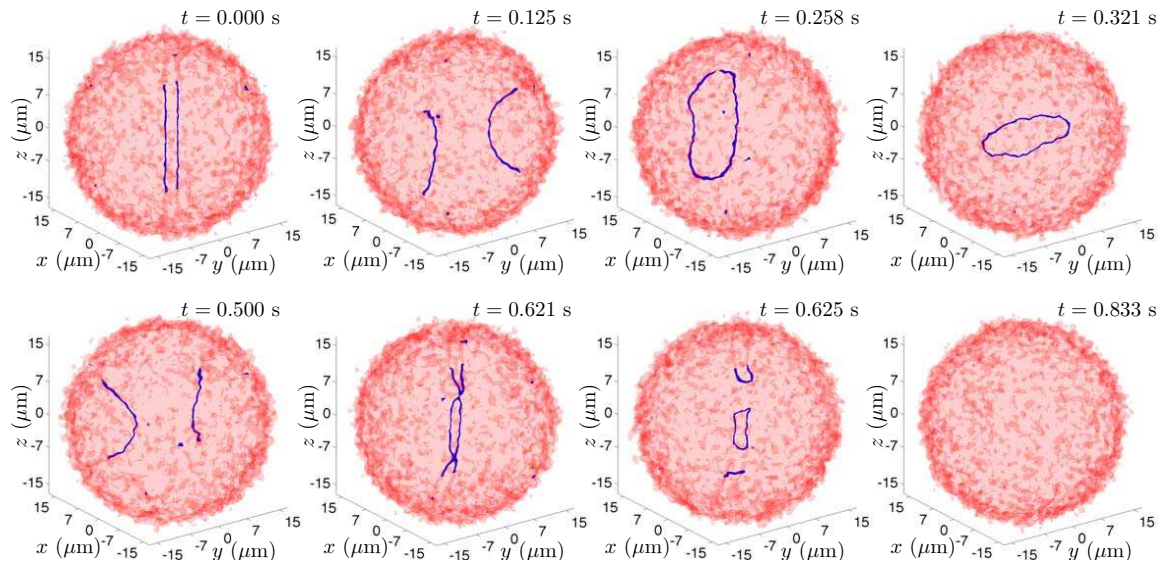}
\caption{Density isosurfaces of the $\rC$-field showing the SPGPE evolution of a vortex dipole with initial vortex locations $(x,y_\pm)=(0,\pm s/2)$, for separation $s=3{\mu \rm m}$. Red: the $\rC$-field isosurface. Blue: isosurface restricted to the region of the vortex cores.}
\label{sphSeq}
\end{center}
\end{figure*}
%%====================================================
In  Fig.~\ref{figvel} we show the decay rates versus ${T_{c({\rm{corr}})}} $   verifying that these systems at the highest temperatures remain below the 3D critical temperature.  However, for oblateness $12\lesssim\Lambda$ we see anomalously rapid increase in the vortex decay rate for temperatures exceeding $\sim 0.95 T_{c({\rm{corr}})}$.  For these cases the system crosses over to being quasi-2D as the chemical potential becomes of the order of the energy to activate degrees of freedom in the tightly confined direction, i.e. $\mu\lesssim\hbar\omega_z+\epsilon_0$ (where $\epsilon_0$ is the single particle ground state energy). In particular, the highest temperature results for $\Lambda=8,12,15,20$ have $\mu/\hbar\omega_z=2.2,1.69,1.42,1.19$  (c.f.~$\hbar\omega_z+\epsilon_0\approx3\hbar\omega_z/2$). In the quasi-2D regime the dynamics of vortices are not well-understood, but strong phase fluctuations will likely facilitate the rate of vortex damping as we observe in our results \footnote{The system still has a phase space density sufficiently high that spontaneous vortices are not expected to be significant (e.g.~see \cite{Hadzibabic2006a,*Simula2006a,*Bisset09d}).}.  A systematic investigation  of vortex decay in a system more deeply in the 2D regime is therefore needed, but is beyond the scope of our work here.

\section{Dipole decay}
Having systematically studied the role of geometry and temperature in single vortex decay, we now consider two particular cases of the decay of a vortex dipole at finite temperature. 

\subsection{Spherical geometry: decay via ring and loop formation}\label{vring}
We first consider dipole decay in a spherical trap. We use SPGPE simulation parameters $(T, \mu, \ecut) = (52~{\rm{nK}}, 37\hbar\bar{\omega},71\hbar\bar{\omega})$, which give a total atom number $N=2\times10^6$ and dimensionless damping rate (\ref{gamdef}) of $\hbar \gamma/ k_B T = 3\times10^{-4}$. These parameters give a system with the same atom number and temperature as the oblate configuration considered in the experiment of Ref.~\cite{Neely10a} and in the next section, so the system serves as a benchmark for the effect of geometry on vortex dipole dynamics in this system. We create a finite-temperature vortex dipole state by imposing a vortex dipole phase pattern on a SPGPE equilibrium state via $\psi_\rC(\x)\to \psi_\rC(\x)e^{i\Theta(\x)}$, where $\Theta(\x)=\arctan{\left(\frac{y-y_1}{x}\right)} - \arctan{\left(\frac{y-y_2}{x}\right)}$.  This creates an axially aligned vortex dipole with positive and negative vortices located at $(x,y_\pm) = (0,\pm s/2)$, with $s=3{\mu \rm m}$. Figure \ref{sphSeq} shows the decay sequence for the vortex dipole. The initially axially aligned vortices rapidly curve as the vortices precess, such that the vortex cores are normal to the surface. This long-wavelength bending leads to formation of a vortex ring, which then subsequently dissociates into two vortices at $t=0.5{\rm s}$. As the vortices closely approach again at $t=0.62{\rm s}$, both loops and rings form, reducing the total vortex length, and leading to total decay in less than 1s. We thus observe rapid vortex dipole decay via intermediate vortex ring and loop formation. 
%====================================================
\begin{figure}[!t]
\begin{center}
\includegraphics[width=\columnwidth]{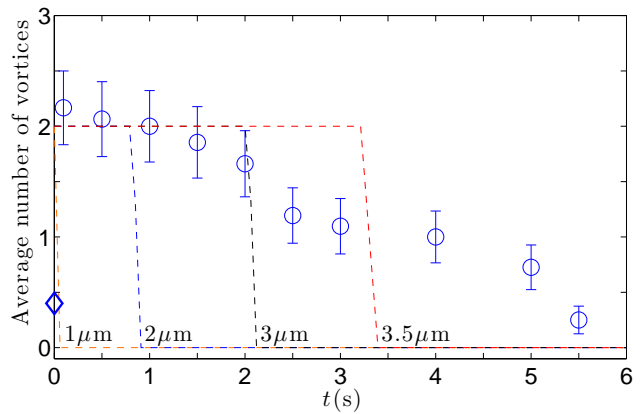}
\caption{Average number of vortices as a function of time during vortex dipole decay for numerical simulations and experimental data. The dashed lines show SPGPE results for a range of initial vortex dipole separations. Circles show experimental data of Ref.~\cite{Neely10a}. The blue diamond at $t = 0$~s shows the average number of vortices observed in the experimental initial state due to spontaneous vortex formation~\cite{Weiler08a}.}
\label{figlifetime}
\end{center}
\end{figure}
%====================================================
%%====================================================
%isosurfaces
\begin{figure*}[!t]
\begin{center}
\includegraphics[width=0.85\textwidth]{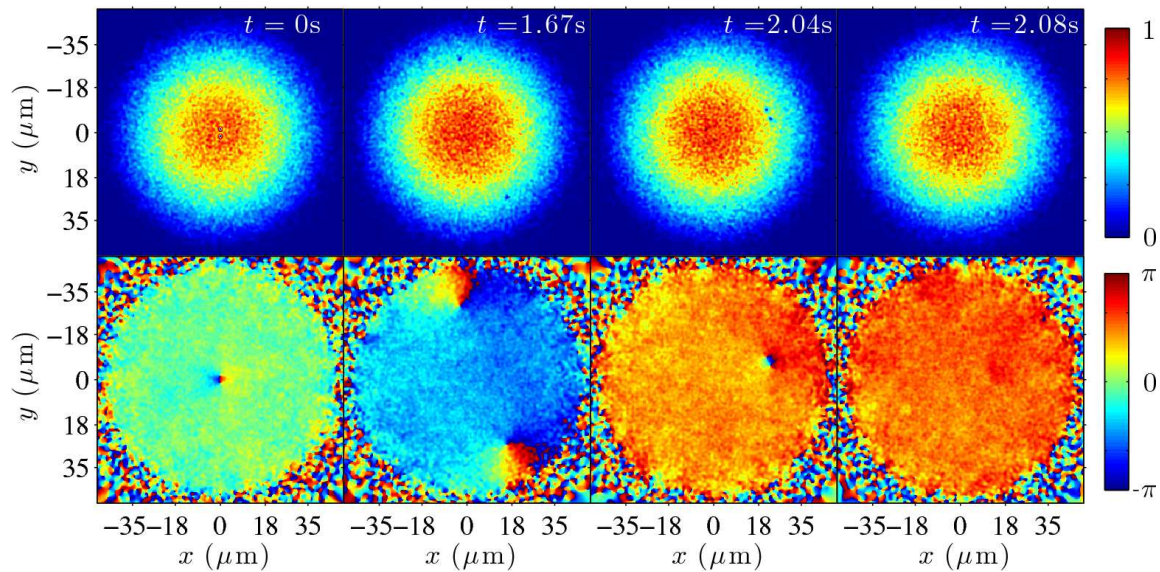}
\caption{Column densities (top) and phase slices for the $\rC$-field showing the SPGPE evolution of a vortex dipole with initial vortex locations $(x,y_\pm) = (0,\pm s/2)$, for separation $s=3~\mu{\rm{m}}$. The series spans two orbits of the dipole, where a single orbital period is $\sim 1.2{\rm s}$, culminating in mutual vortex annihilation towards the completion of the second orbit. The peak density is renormalized to unity at each time (upper colorbar), and the phase is represented on the interval $[-\pi,\pi)$ (lower colorbar).}
\label{figsequence}
\end{center}
\end{figure*}
%%====================================================
\subsection{2DV regime: comparison of experiment and theory}
In this section we use (\ref{SGPEsimp}) to model the dynamics of a charge-one vortex dipole in a highly oblate finite-temperature Bose-Einstein condensate, as in the experiment by Neely {\emph{et al.}}~\cite{Neely10a}.  The system consists of $\sim 2\times 10^6$ atoms at temperature $\sim 52$nK, held in a harmonic trap with frequencies $(\omega_{r},\omega_z) = 2\pi\times(8,90)$Hz. The trap aspect ratio $\Lambda=11.25$ exceeds $\Lambda_c\approx 8$ identified in Sec.~\ref{SecEffOfGeom} for closely related parameters. We can therefore expect that the experimental regime is also approximately that of 2DV. Experimentally, a single vortex dipole is created by pushing the BEC around an past an obstacle above a critical velocity. Ideally, we would like to compare the distance $s$ between the two vortices upon the creation of the dipole in both experiment and simulations, and to relate $s$ to the vortex dipole lifetime in each case. However, the experimental spatial resolution was $\sim 6\mu m$, which is also the approximate vortex separation found immediately after dipole creation in Gross-Pitaevskii simulations of the expriment~\cite{Neely10a}. Since the vortices are not individually resolvable immediately after creation, the experimental data can only place an upper bound on $s$. We are thus unable to carry out quantitative comparison at this point. In this work we instead map out the dependence of dipole lifetime on $s$ by computing the decay dynamics for a range of initial vortex separations.

As in the spherical geometry of Sec.~\ref{vring} we use SPGPE simulation parameters $(T, \mu, \ecut) = (52~{\rm{nK}}, 37\hbar\bar{\omega},71\hbar\bar{\omega})$, which again gives a total atom number $N=2\times10^6$ and dimensionless damping rate (\ref{gamdef}) of $\hbar \gamma/ k_B T = 3\times10^{-4}$.  
As in the spherical case, we phase imprint a vortex dipole into the BEC. We choose the initial separation between vortices as $ s = 1, 2,2.5,3,$ and $3.5~\mu{\rm{m}}$. Note that the condensate radius is $\sim 50\mu{\rm{m}}$, so these separations are small relative to the system size, but larger than the healing length $\xi\approx0.2 \mu\rm m$. 
Qualitatively, our simulations differ from the experiment in several ways: in our simulations we have two vortices initially, a high degree of cylindrical and mirror symmetry, and a well defined total atom number. In the experiment there is occasionally a vortex present from the BEC formation process~\cite{Weiler08a}, an irreducible (albeit very small) cylindrical asymmetry of the harmonic potential, a mirror asymmetry of the stirring potential, and uncertainty in the total atom number. We do not attempt to model all of this complexity here, confining our study to the basic decay process of internal annihilation in the symmetric system, with fixed total atom number~\footnote{Our simulation initial conditions correspond to a grand canonical ensemble for the $\rC$-field region, with well defined average total atom number $N=2\times 10^6=N_{\rI}+N_{\rC}$. We made use of a Hartree-Fock scheme for SPGPE parameter estimation developed in Ref.~\cite{Rooney10a}.}. 

Figure \ref{figlifetime} shows the mean vortex number over time from the experimental data~\cite{Neely10a}, compared with the results of our simulations. In our simulations we always find that the vortices mutually annihilate each other in the center of the BEC, rather than damping at the boundary. The vortices move rapidly when closely approaching each other, nearing the speed of sound $c=0.2 \rm cm/s$. Consequently, most of the orbital time is spent at the outer edge of the BEC, and the time interval in which annihilation may occur is a small fraction of the orbital period ($\sim 10\%$). Thus the vortex number has an abrupt time dependence for a given initial separation, with characteristic timescale given by the orbital period of the vortex dipole, found numerically and experimentally to be $\sim 1.2\rm s$ and independent of $s$ over a wide range~\cite{Neely10a}. In general, the $s$ value selects a given orbit of annihilation, within which there is very small variation of lifetime. 

In detail, we find that for $s=(1,2,2.5,3,3.5)\mu\rm m$ the mean lifetime is $\tau=(0.01,0.86,2.04,2.07,3.30)\rm s$, corresponding to $N_0=(0,1,2,2,3)$ dipole orbits before annihilation. For example, with $s=1\mu{\rm m}$, the dipole self annihilates immediately after creation. The separations $s=2.5\mu\rm m$ and $s=3\mu\rm m$ lead to almost identical lifetimes of about $2\rm s$, corresponding to the significant drop in vortex number seen experimentally for $2{\rm s}\lesssim t \lesssim 2.5{\rm s}$ which we interpret as an experimental indication of the dipole lifetime. 

In Figure \ref{figsequence} we plot a time series of simulation data for $s=3\mu\rm m$ showing internal annihilation at the end of the second orbit. Interestingly, we observe a tilt in the axis of dipole precession caused by thermal fluctuations, which is also observed in experimental absorption images. 

\section{Conclusions and outlook}
Here we summarize our results on single vortex and vortex dipole decay.

\emph{Sub-critical Regime.---}For spherical systems at $T = 0.78T_c^0$, a lone vortex is freely deformed from its initial straight-line configuration by the activation of vortex bending modes.  This effect diminishes as the condensate is flattened, and bending modes become energetically inaccessible. The total bending mode power decreases exponentially as the condensate geometry becomes increasingly two-dimensional.  For geometries with oblateness $\Lambda \geq \Lambda_c (\approx 8)$, the total power asymptotes to zero, indicating bending is strongly inhibited.  The effective vortex decay rate also decreases as $\Lambda$ increases, approaching a constant two-dimensional value for $\Lambda_c\leq\Lambda$. $\Lambda_c$ can therefore be identified as the critical oblateness required for the system to enter the 2D vortex regime.

\emph{Critical Regime.---}The foregoing comments hold for temperatures outside an observed critical regime. When critical fluctuations associated with the phase transition become significant, we find that the vortex decay rate is highly sensitive to temperature and geometry. In particular, we see an anomalously large increase in the decay rate for geometries with $\Lambda_c< \Lambda$, and $0.95 T_{c \rm (corr)}\lesssim T$ that we attribute to phase-fluctuations associated with the system entering the quasi-2D regime.

\emph{Vortex Dipole Decay.---}
We have modeled the experiment of Neely {\em et al}~\cite{Neely10a} which is in the 2DV regime ($\Lambda=11.25$), and found that the approximate time for decay of a vortex dipole in the experiment ($\sim 2.25$s) is consistent with numerical simulations with an initial vortex separation of $s\sim 2.5-3\mu$m. In the simulations with these initial conditions mutual vortex annihilation occurs after two complete orbits, at $t\sim 2$s. In this work we have used $s$ as the fitting parameter; improved experimental resolution would be required for a quantitative comparison without any fitting parameters. In similar computations of spherical ($\Lambda=1$) BECs, the vortices decay much more quickly to rings and other bent geometries that would not be readily visible after only $\sim 100{\rm ms}$ using the experimental methods of Ref.~\cite{Neely10a}. We thus infer from the experiments of Ref. \cite{Neely10a} and the calculations presented here that the experimental system consisting of a BEC with oblateness $\Lambda\approx 11.25$ is within the 2D vortex dynamics regime, consistent with our calculations suggesting a critical oblateness of $\Lambda_c\approx 8$.

%%%%%%%%%%%%%%%%%%%%%%%%%%%%%%%%%%%%%%%%%%%%%%%%%%%

{\bf Acknowledgements:} 
We thank Simon Gardiner for discussions regarding this work.
%\appendix
%******************************************************************************
%\section*{References}

\bibliographystyle{prsty}
%\bibliography{References}

%******************************************************************************

\end{document}